# DYNAMIC LOW-POWER TRAFFIC PATTERN FOR ENERGY-CONSTRAINED WIRELESS SENSOR NETWORKS

Almamoon Alauthman

Department of Electrical Engineering, Al-Balqa Applied University, As Salt, Jordan

## ABSTRACT

*Wireless Sensor Networks (WSNs) are extensively utilized in critical applications, including remote monitoring, target tracking, healthcare systems, industrial automation, and smart control in both residential and industrial settings. One of the primary challenges in these systems is maintaining energy efficiency, given that most sensor nodes rely on limited battery resources. To tackle this problem, this study introduces an energy-saving strategy designed for tree-structured networks with dynamic traffic patterns. The approach focuses on lowering power usage by decreasing the length and occurrence of idle listening— a state where nodes remain active unnecessarily while waiting for data transmissions that may never occur. By reducing this form of energy waste, the proposed approach is designed to extend the operational lifetime and enhance the throughput of the wireless sensor network. Simulation results obtained using the OMNeT++ simulator with the MiXiM framework demonstrate that the solution significantly reduces energy consumption, increases data throughput, and improves overall network efficiency and longevity.*

## KEYWORDS

*WSN, TDMA, TPO, MTPO, Energy Consumption, Sensor Node, Idle Listening, Tree Network*

## 1. INTRODUCTION

Energy efficiency is a critical factor in the data collection process within Wireless Sensor Networks (WSNs). Given that sensor nodes typically operate on limited battery power, minimizing energy consumption is vital for prolonging the network's lifespan [1]. Among the different Medium Access Control (MAC) protocols, Time Division Multiple Access (TDMA) has proven to be a highly effective approach for facilitating energy-efficient data transmission in WSNs [2]. The use of only the small sources of energy on the node leads to poor communication, and interruption of processes and eventual division of the network [3].

WSNs have a wide range of applications, and their development must prioritize several key objectives—chief among them is energy efficiency. A perfect wireless sensor network is expected to operate without consuming a lot of energy besides integrating wise planning. It must ensure the fast and accurate collection of data during significant periods of time, all at the same time having low installation fees and low or minimal maintenance requirements [4]. This is because sensor nodes depend on batteries that have limited lifespans, and replacing them can be costly or even hazardous in sensitive environments such as battlefields. To extend the operational lifetime of WSNs, energy-efficient protocols are essential. TDMA is commonly utilized for this purpose due to its contention-free nature, which minimizes packet loss and retransmissions.

Maximizing energy conservation is crucial for sustaining long-term, continuous data collection in sensor networks [5]. As defined in common practices [6], network lifetime typically refers to the duration until the first sensor node depletes its energy. The design of routing structures significantly affects the number of packets transmitted and received by nodes, and thereby their energy consumption. Efficient routing strategies must be developed to divert traffic away from heavily burdened or energy-depleted nodes, thereby extending overall network lifespan. Moreover, packet collisions represent a significant obstacle to energy efficiency in data collection processes.

This paper presents an enhanced TDMA scheduling scheme aimed at improving energy efficiency and expanding the Wireless Sensor Networks (WSNs) lifetime. A major advantage of the proposed schedule is that it enables the parent node to receive data from its child nodes without experiencing an idle listening state, thereby significantly reducing energy consumption at the node level. To verify the effectiveness of the proposed approach, a mathematical model is developed and compared against the conventional TPO schedule. Furthermore, simulations are conducted to assess and compare the performance of both the original and the modified TPO schemes.

Both analytical and experimental findings demonstrate that the proposed scheduling method offers a substantial improvement in energy efficiency over existing approaches. The remainder of the paper is structured as follows: Section two are provides a review of related work, Section three describes the system model, Section four contain the details of the proposed schedule, Section five highlight the performance analysis with comparison of the proposed schedule with recent study in this field, Section six covers the experimental evaluation, and Section seven concludes the study.

## 2. RELATED WORK

In recent years, TDMA-based scheduling has gained significant attention as an efficient method for collecting sensor data in Wireless Sensor Networks (WSNs). Initial studies concentrated on developing communication schedules that ensure each node interacts with its neighboring nodes once during each cycle [7]. In [8], the authors introduced a slot allocation strategy that leverages only local node information to organize TDMA slots for both data transmission and acknowledgment. By assigning distinct time slots to individual nodes, this method successfully prevents data collisions. Likewise, the RD-TDMA (Randomized Distributed TDMA) protocol, presented in [9], generates efficient scheduling rapidly, aiming to minimize overall network configuration time. In general, TDMA-based medium access protocols surpass contention-based alternatives like CSMA in terms of reliability, channel efficiency, and reduced power usage, particularly in applications with high data rates. Despite the proliferation of TDMA-based protocols aimed at boosting energy efficiency, many face inherent limitations. In low-traffic scenarios, TDMA may lead to inefficient slot usage, whereas contention-based protocols like CSMA are more susceptible to collisions under high traffic loads. The TDMA-based approach in [10] aims to conserve energy but at the cost of reduced throughput. Energy-FDM, a CSMA-based protocol introduced in [11], incorporates transmission power control to lower energy usage, but it experiences high collision rates in dense traffic environments.

TDMA is one of the simplest communication strategies, assigning each node a unique, non-overlapping time slot for transmission. Despite its simplicity, TDMA has received limited attention in the context of Underwater Optical Wireless Communication (UOWC) networks, Moreover, there is a clear shortage of research dedicated to the design of TDMA-based MAC protocols for such environments. This is largely because assigning a fixed slot time to Ad-Hoc nodes that depend on directional links presents significant challenges. A few centralized TDMA-based MAC protocols have been proposed, such as those in [12], where time slot allocation is handled by a sink node or a cluster head (CH). These centralized approaches heavily rely on CHs, which restrict network scalability and is unsuitable for UOWC networks due to their inherently dynamic nature. Consequently, there is a pressing need for a distributed TDMA-based MAC solution. Unfortunately, as noted in [13], no current studies have yet addressed the design of such distributed TDMA-based MAC protocols specifically for UOWC networks. Existing distributed TDMA MAC schemes like those in [14] are typically designed for terrestrial vehicular networks (VANETs) operating in the RF spectrum and are not well-suited for directional UOWC scenarios. In [15], the author introduces the Energy Efficient Mega Cluster Based Routing (EEMCR) protocol, specifically designed for large-scale coverage areas. In [16], an optimized energy efficient downlink VLC system is proposed, leveraging a combination of hybrid non-orthogonal multiple access (NOMA) and reconfigurable intelligent surfaces (RIS). Furthermore, [17] presents a distributed TDMA based MAC protocol named Cluster based Cross layer Multi slot MAC (CCM-MAC), which allocates multiple time slots to nodes dynamically, based on real-time slot occupancy data. The authors [18] concerns the problem of wireless sensor networks (WSN) energy conservation, in cases when communication is realized by the method of virtual multiple-input multiple-output (MIMO).

One such effort is the Traffic Pattern Oblivious (TPO) protocol presented in [19], which offers a TDMA-based MAC design suitable for varying traffic conditions. TPO efficiently manages energy use across different traffic patterns and enables the base station to complete data collection earlier by adapting to traffic loads, thereby enhancing both energy and time efficiency. Building on this, Aram Rasul [20] proposed the extra-bit approach to decrease the number of idle listening states. Each packet carries an extra bit for indicating whether more data will follow, allowing receiving nodes to avoid unnecessary listening and thus conserving energy and reducing latency. The author in [21] proposed approach builds upon the original TPO by delivering greater energy efficiency and enhanced throughput. By minimizing idle listening during data collection, sensor nodes' energy usage approaches the theoretical minimum. In [22], the EA-TDMA (Energy-Aware TDMA) protocol was proposed, focusing on energy-efficient communication between wireless sensor nodes by optimizing time slot allocation and transmission coordination.

## 3. THE SYSTEM MODEL

Assume a WSN that organized in a tree topology, represented by a graph X= (S, V), where S is the set of sensor nodes and V represents the communication links that connect the sensors. The root node, usually a base station or sink, serves as the central hub for data collection. Each sensor node is equipped with a unidirectional transceiver intended to monitor a designated area or application. Communication across the network is carried out on a single frequency channel, and since nodes are half-duplex, they are unable to transmit and receive simultaneously. To manage communication efficiently, a TDMA protocol is employed, dividing time into uniform slots where multiple transmissions can be scheduled concurrently. This TDMA-based MAC protocol helps avoid packet collisions and unnecessary idle listening, thereby enhancing energy efficiency. The primary objective is to collect data from all sensors using a scheduling approach that minimizes the total number of slots in the TDMA scheduling. By reducing idle listening states, the approach significantly decreases energy consumption, thereby extending the network's overall lifetime. It is important to recognize that not every node generates data in every collection round. If a transmission from any child sensor node to its parent sensor node is scheduled in a default TDMA slot but the child node has no data to send, making an idle listening state, energy waste from idle listening can be avoided. In these cases, both nodes N and P can turn off their transceivers during that slot, conserving energy and potentially reducing power consumption to zero. To maintain schedule validity, each node must transmit its sensed data along with any data received from its child nodes. However, a node cannot transmit more data than the total amount it has generated or collected from its subtree, ensuring that data forwarding remains energy-efficient and logically consistent. To effectively assess the energy efficiency of the proposed Enhanced Traffic Pattern Oblivious (ETPO) scheduling scheme, a standard radio energy model commonly employed in previous wireless sensor network studies is utilized as shown in table 1, These parameters were integrated into the Castalia/OMNeT++ simulation environment to accurately reflect physical-layer characteristics observed in practical deployments. To maintain fairness and ensure the reproducibility of results, the same energy model was consistently applied across all comparisons involving ETPO, MTPO, and TPO.

## 4. PROPOSED SCHEDULED (ENHANCED TRAFFIC PATTERN OBLIVIOUS)

Wenbo Zhao and Xueyan Tang introduced the Traffic Pattern Oblivious (TPO) scheduling approach, also referred to as successive slots scheduling. This method requires that all transmissions from a node to its parent take place in consecutive TDMA slots, beginning with the first slot assigned to that node. No empty slots are allowed between transmissions. This design reduces idle listening, which happens when a parent node waits for data from a child node that has no data to send. For example, if a node has 20 children but only 9 need to transmit data, the TDMA schedule should allocate just 10 consecutive slots without any interruptions.

One of the key advantages of TPO scheduling is its predictability: when a parent node encounters an empty time slot, it can be certain that the corresponding child has no more data to send in the current or upcoming cycles. This enables the parent to switch off its transceiver, reducing energy consumption by eliminating unnecessary idle listening. In the TPO model, data collection begins at the leaf nodes and moves upward toward the base station. The parent node monitors its children's transmissions until an idle slot signals that a particular child is finished sending data. Once the parent has received at least one packet from each child, it aggregates the data and transmits it to its own parent. This process repeats up the hierarchy until the base station gathers all the data and can cease listening. The proposed method builds on the TPO concept by introducing optimizations to eliminate idle listening at intermediate levels of the tree of height n.

- **At the leaf level (Level N):**
  Using the TPO method, each parent node allocates a number of TDMA slots corresponding to the number of its children. However, some idle listening can still occur, as the parent must wait for an empty slot to verify that a child has finished transmitting.
- **At Upper level N-1:**
  When an empty TDMA slot is obtained, the parent node recognizes that the child has completed its data transmission for the current and upcoming rounds. It then switches off its transceiver to save energy and records the total number of packets received.
- **Upward Reporting:**
  Parents at Level N-1 send a message to their parents at Level N-2, informing them of the exact number of packets received. This allows upper-level parents to anticipate incoming traffic.
- **At Level N-2 and higher (up to the base station):**
  Each parent uses the information from its children to create a TDMA schedule with precisely the number of slots needed. Once it receives all expected packets, it immediately turns off its transceiver, avoiding any idle listening.

The Key Distinction from the proposed approach with Standard TPO is that, in conventional TPO, parents listen for one extra slot beyond the number of packets to confirm the end of transmissions. In contrast, the proposed method restricts this extra slot only to parents at the leaf level. At all higher levels, idle listening is eliminated because parents know in advance how many packets to expect, enabling exact TDMA scheduling and more efficient transceiver usage. The Pseudocode of our proposed ETPO is shown in the figure 1.

```
Initialize:
    For each node n ∈ N:
     S [n] ← null
Define Assign (Sensor):
1. Idle_Slots ← 0
    2. For each sensor neighbor ∈ Sensor_Neighbors(Sensor):
      if Time [Neighbor] ≠ null:
         Add Sensor[Neighbor] to Idle listen
    3. Slot ← 0
    4. While Slot ∈ Idle Listen :
      Slot ← Next Slot
    5. Assign[Sensor] ← Next Slot
    6. For all children(Sensor):
      Assign_Slot(Child)
Repeat_Slot(Sensor)
Reurn Assign (Sensor)
```

Figure 1. The Pseudocode for the proposed approach (ETPO)

### 4.1. Lifetime Definitions

The concept of lifetime in Wireless Sensor Networks (WSNs) has been interpreted in multiple ways across the literature. Broadly, it represents the time span during which the network remains operational and can effectively carry out its designated functions [23]. One widely used definition considers network lifetime as the time until the first sensor node—or a set of nodes—exhausts its energy and ceases to function [24]. Another perspective defines it more precisely as the point when the very first node in the network runs out of power [25]. Alternatively, it may be viewed as the longest duration over which the deployed sensors can continuously observe the target phenomenon [26]. In essence, network lifetime is a time-based metric that reflects the functional longevity of sensor nodes.

Energy efficiency, initially described as the ratio of total data that is successfully be transmitted to the total energy consumed [27], plays a crucial role in determining network longevity. The higher the amount of data delivered per unit of energy, the greater the network's energy efficiency. Since a node's lifetime is governed by its energy usage, minimizing energy consumption directly contributes to extending the network's lifespan. In WSN applications, a node's energy consumption typically includes energy used for transmitting, receiving, idle listening, data sampling, and sleep modes. Reducing energy use in these activities is essential for maximizing network lifetime.

$$E = E_{tx} + E_{rx} + E_{Idle} + E_{sleep} \qquad (1)$$

In Equation (1), ETx, ERx, EIdle, and Esleep represent the energy consumed by a node during transmission, reception, idle listening, and sleep states, respectively [28]. As demonstrated, reducing EIdle leads to a decrease in the overall energy consumption. Lower energy usage extends the lifespan of individual sensor nodes, which in turn contributes to maximizing the overall lifetime of the entire Wireless Sensor Network (WSN).

## 4.2. Number of Idle Listening

In the TPO scheduling approach, an idle state occurs when a child node tries to transmit to its parent node, but at least one node within its subtree has no data to send. Consequently, the total occurrences of idle state correspond to the number of nodes that have no data to transmit. By contrast, the proposed approach limits idle listening to transmissions at the tree's last level, and only when none of the nodes at that level possess data. As a result, idle listening in this approach is limited to instances where nodes at the lowest level have no data to transmit, significantly reducing the number of idle listening occurrences compared to the TPO method. For instance, in the tree shown in Figure 1, where only nodes C, D, E, and F contain data, the TPO scheme results in four idle listening occurrences: during transmissions from C to A, F to B, A to the sink R, and B to R. Meanwhile, the proposed method reduces idle listening to just two instances—during transmissions from C to A and F to B—showing a clear improvement in efficiency.

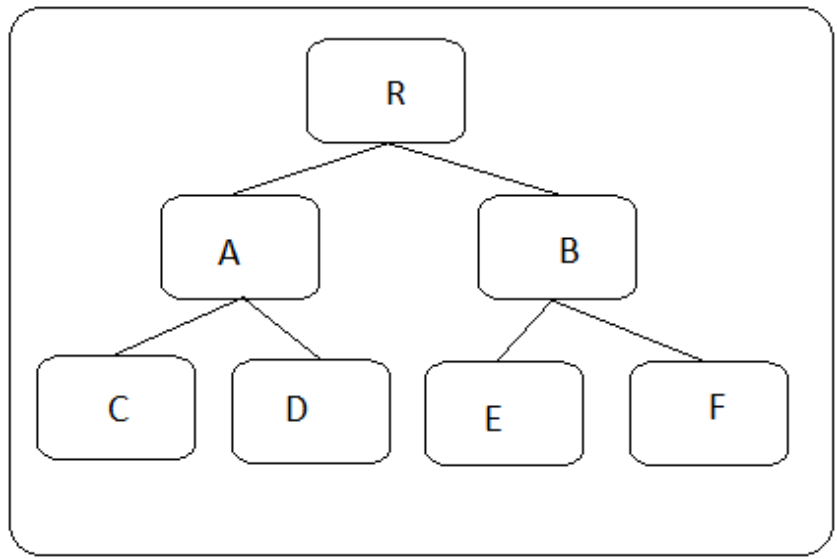

Figure 2. Tree network with only (C, D, E, F) have a data

To evaluate the idle listening state occurrences in both the TPO technique and the proposed scheduling, a probabilistic model is employed, each node has a probability P (ranging from 0 to 1) of having data to transmit. When P = 0, it indicates that none of the nodes in the tree have data to send, whereas P = 1 means that all nodes are actively transmitting data. Both techniques result in an equal number of idle listening events only in two specific scenarios: when P=0, since no data exists at any node, and when P=1, because all nodes are actively transmitting, resulting in zero idle listening for both. For intermediate values of P between 0 and 1, however, the TPO technique experiences more idle listening events than the proposed method, demonstrating that the proposed approach more effectively reduces idle listening in typical scenarios.

### 4.2.1. Balanced Tree

A balanced tree is defined as a structure in which every parent node has the same number of child nodes, represented by R. When applying the TPO technique, the total number of idle listening instances in this balanced tree can be determined using the following calculation:

$$TPO\ (Idle\ state\ number) = (R^{L+1} - 1) \times (1 - P) \qquad (2)$$

On the other hand, when applying the proposed technique, the idle listening count in a balanced tree is:

$$Proposed\ (Idle\ state\ number) = R^L \times (1 - P) \qquad (3)$$

In these formulas, P is the probability that any given sensor has data to transmit, L is the height of the tree, and R is the number of children for each parent node.

### 4.2.2. Unbalanced Tree

An unbalanced tree is defined by an unequal distribution of children across parent nodes, indicating that the value of R varies throughout the structure. To estimate the number of idle listening events in such a tree, let s represent the size of a subtree, measured from the lowest level up to—but excluding—the sink node. Let P be the probability that a sensor node has data to transmit, and L is the number of subtrees of size K. In the context of the TPO approach, idle listening occurs if at least one node within a subtree lacks data. As a result, the total number of idle listening instances can be calculated as:

$$TPO\ (Idle\ state\ number) = \sum_{L=0}^{N} L(1 - P^K) \qquad (4)$$

In the proposed approach, idle listening takes place during a child-to-parent transmission only when all nodes within the associated subtree lack data to transmit. Thus, the total number of idle listening events can be represented as:

$$Proposed\ (Idle\ state\ number) = \sum_{L=0}^{N-1} L(1 - P)^K \qquad (5)$$

Consider the following example for unbalanced network as shown in figure 3

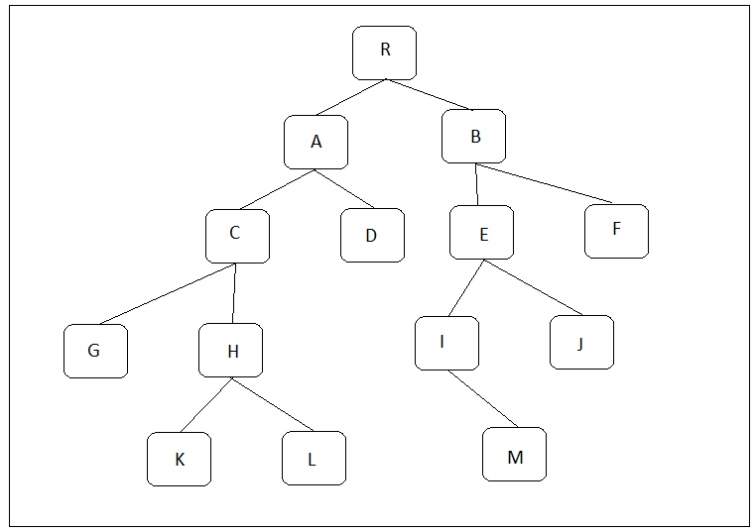

Figure 3. Unbalanced tree of height 4

To compute the number of idle listening using equation (4) for TPO, and equation (5) for proposed technique, the tree shown in figure 3 have a 7 sub tree with size 1, 3 sub tree with size 3, one sub tree of size 5, one sub tree with size 7, and there is no any sub tree with size 2,4,6 The number of idle listening using TPO technique is equal 8 idle listening, and for the proposed technique is equal 4 idle listening state.

International Journal of Computer Networks & Communications (IJCNC) Vol.17, No.5, September 2025

## 5. SIMULATION SETUP

This section outlines the simulation parameters used to highlight the performance of the proposed scheduling approach using the Omnet++ simulation platform. To demonstrate its effectiveness in enhancing network lifetime and throughput, the proposed method is evaluated against the original TPO and MTPO approaches. The detailed parameters utilized in these simulations are listed in Table 1.

Table 1: Simulation Setup

| Number of node | Varying from 32 to 1024 |
|---|---|
| Beginning energy level | 5 Joul |
| Simulation Time | 150 sec. |
| Energy consumption during data aggregation | 25nj\bit |
| The radio module depletes its energy | 25nj\bit |
| Area (m^2) | 50m x 50m |
| Packet size | 1024 bit |

Finally, graphs showing Energy Consumption versus Number of Sensors and Throughput versus Number of Sensors were generated. The simulations were carried out using OMNeT++ version 4.6 with the MiXiM framework.

### 5.1. Energy Consumption

To evaluate the energy efficiency of the ETPO model, two simulation scenarios were performed. In the first scenario, the number of nodes varied from 10 to 1100 in increments of 200. The simulation ran until 150 data packets were successfully received, after which the average energy consumed to deliver these packets was measured. ETPO's energy performance was then compared with other models, as shown in Figure 4. The results reveal that ETPO consumes at least 2% less energy than MTPO and up to 40% less than TPO when the network has 100 nodes. As the node count grows to 1100, ETPO achieves a 31% reduction in energy use compared to MTPO and an impressive 91% reduction relative to TPO.

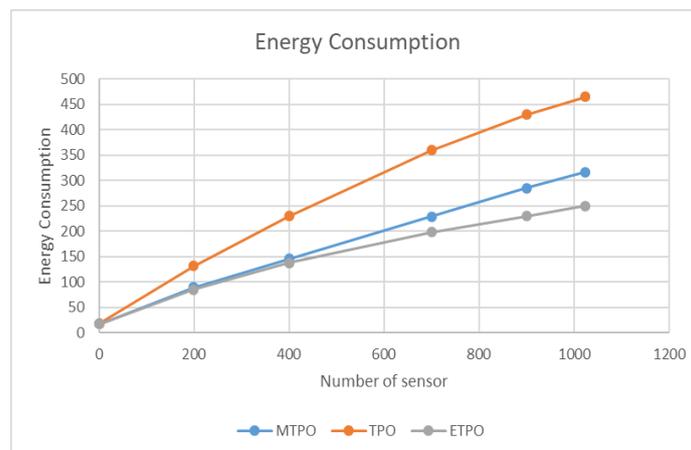

Figure 4. Power consumption using MTPO, TPO, and the proposed technique ETPO

International Journal of Computer Networks & Communications (IJCNC) Vol.17, No.5, September 2025

Table 2 presents the energy consumption results for a small-scale network consisting of up to 32 sensor nodes, with the number of sensors varying from 2 to 32. The simulation was run until 10 data packets were successfully transmitted, after which the average energy used for their delivery was measured. As indicated in the table, the proposed ETPO approach demonstrates significantly lower energy consumption compared to both TPO and MTPO. This improvement suggests that ETPO contributes to extending the overall network lifetime, aligning with the primary objective of this study.

Table 2. Energy Consumption Comparison among TPO, MTPO, and ETPO

|  | Energy Consumption | | |
| --- | --- | --- | --- |
| Number of nodes | TPO | MTPO | ETPO |
| 2 | 22 | 20 | 17 |
| 4 | 23 | 21 | 18 |
| 8 | 27 | 22 | 19 |
| 12 | 29 | 23 | 20 |
| 16 | 31 | 24 | 22 |
| 20 | 34 | 26 | 23 |
| 26 | 37 | 30 | 25 |
| 32 | 43 | 35 | 27 |

Figure 5 illustrates the objective of the third simulation, providing a comparison of energy consumption across network nodes using TPO, MTPO, and the proposed ETPO method. The simulation highlights differences in residual energy levels among the three approaches. Specifically, the TPO curve maintains linearity for approximately 950 rounds, MTPO extends this to 1600 rounds, while the proposed scheduling approach ETPO sustains linearity up to 4850 rounds. The total remaining energy in TPO declines to nearly zero after approximately 3550 iterations, while MTPO retains energy up to 5500 iterations. In contrast, ETPO extends node energy usage up to 5100 iterations. These results confirm that ETPO surpasses traditional methods by delivering better network longevity, higher throughput, and extended overall lifespan. Consequently, the ETPO approach more efficiently Expand the lifetime of wireless sensor networks compared to both TPO and MTPO Approaches.

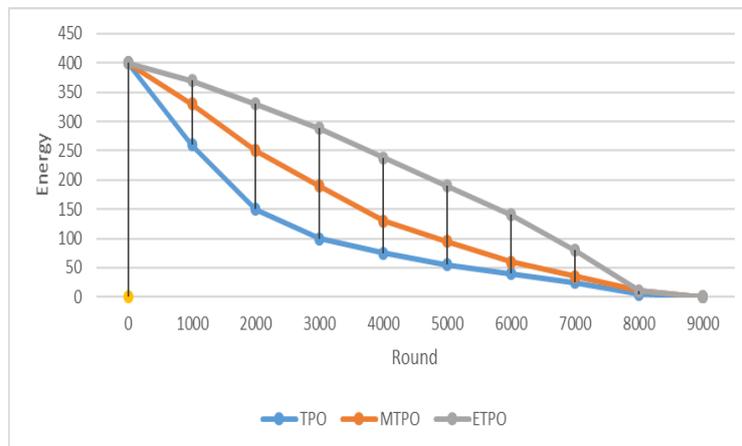

Figure 5. Total energy consumption.

International Journal of Computer Networks & Communications (IJCNC) Vol.17, No.5, September 2025

## 5.2. Throughput

Throughput is a crucial measure of network performance, representing the total number of data packets that successfully sends to the destination node (Sink) within a given time period. Generally, throughput rises as the number of nodes grows, due to a larger volume of data being transmitted throughout the network. While individual nodes may send varying amounts of data, the overall increase in throughput reflects enhanced network efficiency.

Figure 6 illustrates the throughput performance of TPO, MTPO, and the proposed ETPO approach across varying numbers of sensors from 1 to 1024 in the network over 1000 rounds. The results clearly show that the ETPO approach achieves higher throughput compared to both TPO and MTPO.

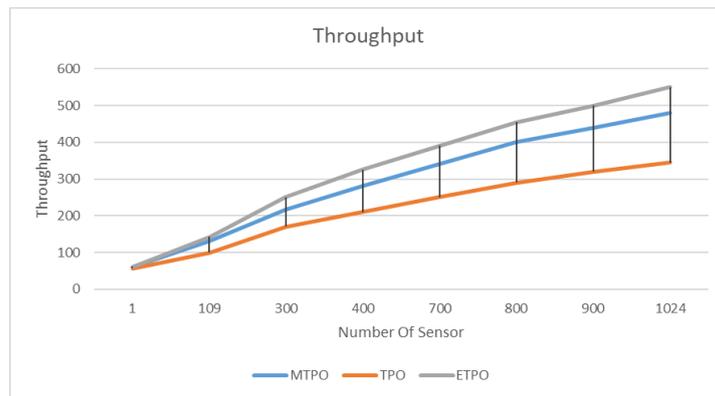

Figure 6. Network Throughput using TPO, MTPO and ETPO

Figure 7 illustrates the throughput performance of a network with 400 sensor nodes across different rounds. The TPO approach shows the lowest throughput, mainly due to the majority of sensors exhausting their energy and ceasing operation after 500 rounds. Conversely, both MTPO and ETPO keep sensors active throughout all rounds, with ETPO achieving higher throughput than MTPO, indicating greater efficiency.

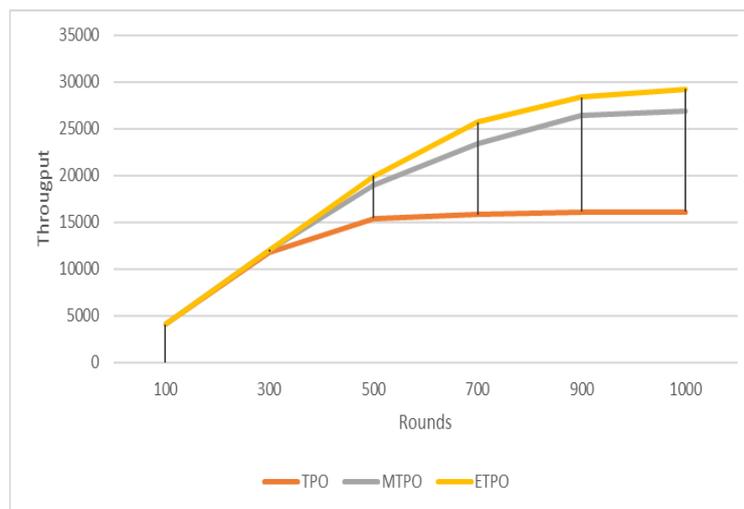

Figure 7. Throughput for the TPO, MTPO, and the ETPO technique

## 5.3. Number of Dead Sensor

As described in [29], network lifetime refers to the duration during which the deployed sensor nodes remain active and continue monitoring the target phenomenon. A typical method to assess this is by counting the time of the first sensor node that depletes its energy and being dead. This is an important metric for determining how long the network maintains its sensing functions [30]. As illustrated in Figure 8, the count of inactive nodes differs across various strategies: all nodes in the TPO-based network are depleted by round 5800, in MTPO by round 6500, while the proposed scheduling approach ETPO prolongs operation until round 7000. These results clearly demonstrate that ETPO significantly enhances network lifetime, aligning with the primary goal of this study.

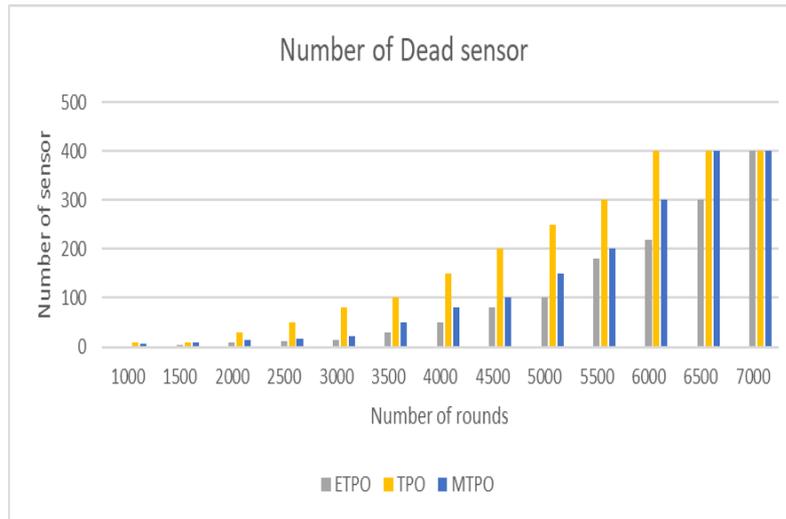

Figure 8. A comparison of dead sensors using TPO, MTPO, and ETPO technique

In the next simulation, the time between starting the network to collect data, until the first sensor became death is computed. the number of sensors is varied between 100 and 1000. As shown in Figure 9. The time until the first sensor is dead is computed for a network environment with 100 sensors. It takes 9.1s for the TPO technique to die, 14s for MTPO, and 19s for ETPO technique.

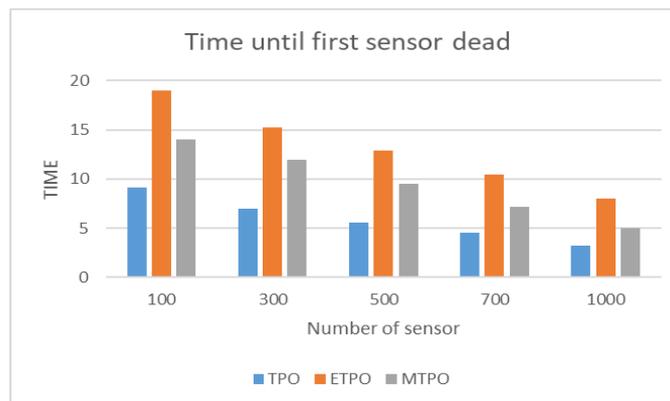

Figure 9. Time until first Sensor Dead using TPO, MTPO, and ETPO Technique

## 6. CONCLUSION

This paper introduces an enhanced TDMA-based scheduling method named Enhanced Traffic Pattern Oblivious (ETPO), which builds upon the original Traffic Pattern Oblivious (TPO) approach. The primary objective is to extend the operational lifetime and enhance the throughput of wireless sensor networks (WSNs). The improved Dynamic Low-Power Traffic Pattern mechanism, integrated within the TDMA structure, delivers substantial gains in energy efficiency, network longevity, and data throughput for energy-constrained WSNs. By effectively reducing idle listening during data collection phases, the energy consumption of sensor nodes approaches optimal levels. A theoretical comparison with the baseline TPO method, along with simulation experiments conducted using the MiXim/OMNeT++ platform, highlights the benefits of ETPO. The results demonstrate a significant decrease in energy consumption and an increase in data throughput when compared not only to TPO but also to more recent scheduling techniques in the field. While TPO, MTPO approaches maintain network functionality for approximately 700–800 rounds, ETPO extends this duration to over 1,150 rounds—marking a 30–40% improvement in network lifetime. Furthermore, it reduces energy usage per transmission cycle by up to 32%. In terms of data delivery, ETPO improves throughput by 17–22%, particularly under conditions involving irregular or bursty traffic patterns. These findings validate the robustness and efficiency of the proposed scheduling strategy, positioning ETPO as a highly competitive solution for modern, delay-sensitive, and energy-critical WSN applications.

### CONFLICT OF INTEREST

The authors declare no conflict of interest.

### REFERENCES

[1] Chen, J., Chang, Y., and Wen, L., (2023). A Long-Distance First Matching Algorithm for Charging Scheduling in Wireless Rechargeable Sensor Networks. Energies, volume:16, issue:18, DOI: 10.3390/en16186463

[2] Guan, S., Zhang, J., Song, Z., Zhao, B., and Li, Y., (2020). Energy-Saving Link Scheduling in Energy Harvesting Wireless Multihop Networks With the Non-Ideal Battery. in IEEE Access, volume: 8, page: 144027-144038, DOI: 10.1109/ACCESS.2020.3014645

[3] Cong, H., & Ho T., (2016). Energy Savings in Applications for Wireless Sensor Networks Time Critical Requirements. International Journal of Computer Networks and Communications, vol: 8, issue: 4, pp: 47-58, DOI: 10.5121/ijcnc.2016.8403

[4] Gholamreza Farahani (2019). Energy Consumption Reduction in Wireless Sensor Network Based on Clustering. International Journal of Computer Networks & Communications (IJCNC), Vol: 11, No: 2, pp:33-51.

[5] Ntabeni, U., Basutli, B., Alves, H and Chuma, J., (2024). Device-Level Energy Efficient Strategies in Machine Type Communications: Power, Processing, Sensing, and RF Perspectives. in IEEE Open Journal of the Communications Society, volume: 5, page: 5054-5087, DOI: 10.1109/OJCOMS.2024.3443920.

[6] W. Zhao, Y. Li, B. Yan and L. Xu, (2020). Median Access Control Protocols for Sensor Data Collection: A Review. in IEEE Access, volume 8, page: 160078-160098, DOI: 10.1109/ACCESS.2020.3019392.

[7] Pushpa .S, Mangayarkarasi, P., (2013) An Energy Efficient MAC Protocol For Data Collection With Dynamic Traffic Pattern In Wireless Sensor Network. International Journal of Engineering Research & Technology (IJERT), Volume: 2, Issue: 6.

[8] Guglielmo, D., Anastasi, D., and Conti, M., (2013). A localized slot allocation algorithm for wireless sensor networks. 2013 12th Annual Mediterranean Ad Hoc Networking Workshop (MED-HOC-NET), page: 89-96, DOI: 10.1109/MedHocNet.2013.6767415.

[9] Ashutosh Bhatia, R.C. Hansdah, (2016). TRM-MAC: A TDMA-based reliable multicast MAC protocol for WSNs with flexibility to trade-off between latency and reliability, Computer Networks, Volume 104, Pages 79-93, DOI: https://doi.org/10.1016/j.comnet.2016.04.018.

[10] Chun-Cheng, L., Teng-Huei, C., and Hui-Hsin, C, (2016). Adaptive router node placement with gateway positions and QoS constraints in dynamic wireless mesh networks. Journal of Network and Computer Applications, volume: 17, issue: 2 DOI: 10.1016/j.jnca.2016.05.005

[11] Xin, Y., Ling, W., Jian, X., Zhaolin, Z., and Juan F., (2018). Energy Efficiency TDMA/CSMA Hybrid Protocol with Power Control for WSN. Wireless. Communication. Mobile Computers. Volume: 12, issue: 3, DOI: https://doi.org/10.1155/2018/4168354

[12] Ji Huang, Yinghao Lu, Zhenlong Xiao, Xin Wang, (2018). A novel distributed multi-slot TDMA-based MAC protocol for LED-based UOWC networks, Journal of Network and Computer Applications, Volume 218, issue: 2, DOI: https://doi.org/10.1016/j.jnca.2023.103703.

[13] Fang, C., Li, S., Wang, Y., & Wang, K. (2023). High-Speed Underwater Optical Wireless Communication with Advanced Signal Processing Methods Survey. Photonics, volume: 10, issue: 7, DOI: https://doi.org/10.3390/photonics10070811

[14] Hota, L., Nayak, B. P., Kumar, A., Ali, G. G. M. N., & Chong, P. H. J. (2021). An Analysis on Contemporary MAC Layer Protocols in Vehicular Networks: State-of-the-Art and Future Directions. Future Internet, volume: 13, issue: 11, DOI:. https://doi.org/10.3390/fi13110287

[15] Vellaichamy, J., Basheer, S., Bai, P. S. M., Khan, M., Kumar Mathivanan, S., Jayagopal, P., & Babu, J. C. (2023). Wireless Sensor Networks Based on Multi-Criteria Clustering and Optimal Bio-Inspired Algorithm for Energy-Efficient Routing. Applied Sciences, volume: 13, issue: 5, DOI: https://doi.org/10.3390/app13052801

[16] A. Ihsan, M. Asif, W. U. Khan, I. N. O. Osahon and S. Rajbhandari, (2024). Energy-Efficient TDMA-NOMA for RIS-Assisted Ultra-Dense VLC Networks. in IEEE Transactions on Green Communications and Networking, DOI: 10.1109/TGCN.2024.3511114.

[17] Ji Huang, Yinghao Lu, Zhenlong Xiao, Xin Wang, (2023) A novel distributed multi-slot TDMA-based MAC protocol for LED-based UOWC networks, Journal of Network and Computer Applications, Volume 218, DOI: https://doi.org/10.1016/j.jnca.2023.103703.

[18] Shitiz U., & Mahaveer S., (2025). Energy Efficient Virtual MIMO Communication Designed for Cluster based on Cooperative WSN using Oleach Protocol. International Journal of Computer Networks & Communications (IJCNC) Vol: 17, No: 3, pp: 74-88

[19] Zhao, W., Tang, x., (2013). Scheduling sensor data collection with dynamic traffic patterns. IEEE Trans. Parallel Distributed System, volume 24, issue 4, page: 789–802, DOI: https://ieeexplore.ieee.org/document/6205748

[20] Rasul, A., Erlebach, T., (2018). The extra-bit technique for reducing idle listening in data collection. International Journal of Sensor Networks (IJSNET), volume: 25, issue: 1, DOI: 10.1504/IJSNET.2017.086788

[21] Alauthman A., Nik W.N.S.W., Mahiddin N.A. (2021). An Adaptive Low Power Schedule for Wireless Sensor Network. IT Convergence and Security. Lecture Notes in Electrical Engineering (LNEE), Springer, volume: 712, DOI: 10.1007/978-981-15-9354-3_17

[22] Rathna, R., Mary, L., Sybi, J., and V Maria, A., (2021). Energy Efficient Cross Layer MAC Protocol for Wireless Sensor Networks in Remote Area Monitoring Applications. Journal of Information Systems and Telecommunication (JIST), volume35, issue 9, page: 207-217, DOI: https://www.jist.ir/Article/15479/FullText

[23] Rama, V., Vuppala, S., Mohd, A, (2024). Enhancing Wireless Sensor Network lifetime through hierarchical chain-based routing and horizontal network partitioning techniques. Measurement: Sensors, Volume 36, DOI: https://doi.org/10.1016/j.measen.2024.101300.

[24] Meenakshi, Y., Anoop, B., Jha, C., (2017). Low –Energy Adaptive Clustering Hierarchy using Virtualgrid Method. International Journal of Advanced Research in Computer Science, volume : 8, issue: 8, DOI: http://dx.doi.org/10.26483/ijarcs.v8i8.4702

[25] Bekal P, Kumar P, Mane PR, Prabhu G. (2024). A comprehensive review of energy efficient routing protocols for query driven wireless sensor networks. F1000Res. Vol: 4, issue: 12, DOI: 10.12688/f1000research.133874.2.

[26] Sanjoy, M., Saurav, G., Sunirmal K., (2025). Energy-efficient data reduction and reconstruction schemes to enhance network lifetime in wireless sensor networks. International Journal of Sensor Networks volume:47, issue:2, page: 72-87, DOI: 10.1504/IJSNET.2025.144564

[27] Hatem, A., Lars, H. (2024) Parametric Machine Learning-Based Adaptive Sampling Algorithm for Efficient IoT Data Collection in Environmental Monitoring. Journal of Network and Systems Management. Journal of Network and Systems Management, volume: 33, issue 1, DOI: 10.1007/s10922-024-09881-1

[28] Ting, X., Ming, Z., Xin, Y., and Yusen, Z., (2022). An improved communication resource allocation strategy for wireless networks based on deep reinforcement learning, Computer Communications, Volume 188, Pages 90-98, DOI: https://doi.org/10.1016/j.comcom.2022.02.018.

[29] Takabayashi, K., Tanaka, H., Sugimoto, C., Sakakibara, K., & Kohno, R. (2018). Performance Evaluation of a Quality of Service Control Scheme in Multi-Hop WBAN Based on IEEE 802.15.6. Sensors, volume:18, issue:11, https://doi.org/10.3390/s18113969

[30] Yoon, C., Cho, S., & Lee, Y. (2024). Extending WSN Lifetime with Enhanced LEACH Protocol in Autonomous Vehicle Using Improved K-Means and Advanced Cluster Configuration Algorithms. Applied Sciences, volume: 14, issue: 24, DOI: https://doi.org/10.3390/app142411720

**AUTHOR**

**Almamoon Alauthman** earned his Bachelor's degree in Computer Engineering from Al-Balqa Applied University (BAU) in Amman, Jordan, and his Master's degree in Computer Engineering from Jordan University of Science and Technology (JUST) in Irbid, Jordan. He completed his Ph.D. at Sultan Zainal Abidin University (UniSA) in Malaysia. Currently, he serves as an assistant professor in the Electrical Engineering Department at the Faculty of Engineering, Al-Balqa Applied University. His research interests encompass VLSI design, parallel processing, neural networks, computer architecture and organization, and wireless networks.

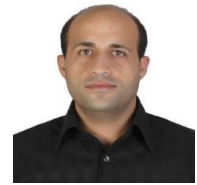